\documentclass[useAMS,usenatbib,fleqn]{mn2e}

\usepackage{graphicx}
\usepackage{textcomp}
\usepackage{amsmath}
\usepackage{breqn}
\usepackage{wasysym}
\usepackage{amssymb}
\title[Effects of refraction on transmission spectra of gas giants]
{Effects of refraction on transmission spectra of gas giants: decrease of the 
Rayleigh scattering slope and breaking of retrieval degeneracies}
\author[Yan B\'etr\'emieux]{Yan B\'etr\'emieux$^{1}$ \thanks{personal e-mail: yanbet@yahoo.com}\\
$^{1}$Max-Planck-Institut f\"{u}r Astronomie, K\"{o}nigstuhl 17, D-69117 Heidelberg, Germany}

\begin{document}


\pagerange{\pageref{firstpage}--\pageref{lastpage}} \pubyear{2015}

\maketitle

\label{firstpage}

\begin{abstract}

Detection of the signature of Rayleigh scattering in the transmission spectrum of an exoplanet is increasingly 
becoming the target of observational campaigns because the spectral slope of the Rayleigh continuum enables one 
to determine the scaleheight of its atmosphere in the absence of hazes. However, this is only true when one ignores 
the refractive effects of the exoplanet's atmosphere. I illustrate with a suite of simple isothermal clear Jovian 
H$_2$-He atmosphere models with various abundances of water that refraction can decrease significantly the spectral 
slope of the Rayleigh continuum and that it becomes flat in the infrared. This mimics a surface, or an optically thick 
cloud deck, at much smaller pressures than one can probe in the non-refractive case. 
The relative impact of refraction on an exoplanet's transmission spectrum decreases with 
atmospheric temperatures and increases with stellar temperature. Refraction is quite important 
from a retrieval's perspective for Jovian-like planets 
even at the highest atmospheric temperatures (1200~K) 
considered in this paper, and for all stellar spectral
types.
Indeed, refraction breaks in large part the retrieval degeneracy 
between abundances of chemical species and the planet's radius because the size of spectral features 
increases significantly with abundances, in stark contrast with the non-refractive case which simply shifts them 
to a larger or smaller effective radius. Abundances inferred assuming the atmosphere is cloud-free are lower limits. 
These results show how important it is to include refraction in retrieval algorithms to 
interpret transmission spectra of gas giants accurately.

\end{abstract}


\section{Introduction}\label{intro}

Transmission spectroscopy is a powerful observational method that allows one to probe the atmosphere of a transiting 
exoplanet. Occultation by the body of the planet and absorption by the planetary atmosphere cause a drop in the 
stellar flux during the planetary transit. The absorption cross section of the exoplanet as a function of wavelength, 
expressed in terms of its effective radius \citep{Brown_2001}, depends on the transmission of the atmosphere at the terminator
as a function of altitude, as well as the radius of the planet at a reference atmospheric pressure. 

To understand which atmospheric parameters can be retrieved from the analysis of transmission spectra, 
\citet{Lecavelier_2008} derived a few simple relationships for an isothermal atmosphere with a uniform 
vertical composition. They showed numerically that the effective radius of an exoplanet occurs at an altitude where 
the integrated optical depth along the line of sight grazing the planetary limb, or slant optical depth, is about 0.56. 
This was later demonstrated analytically by \citet{dW_S_2013}. Using this simple relation, \citet{Lecavelier_2008} also 
showed that the size of spectral features scales with the atmospheric scaleheight. Assuming H$_2$ Rayleigh scattering to 
be the most important source of extinction from 0.6 to 1~\micron, they derived the scaleheight, and hence the temperature,
of HD~189733b from transit data obtained with the Hubble Space Telescope \citep{Pont_2008}. 

Since the spectral dependence of Rayleigh scattering cross sections is well known, spectral regions dominated by
Rayleigh scattering are particularly useful to determine the scaleheight of an atmosphere by measuring the 
change in the effective planetary radius with wavelength (\citealt{Lecavelier_2008}; \citealt{B_S_2012}; \citealt{H_B_2012}). 
This Rayleigh slope can be used to discriminate between various bulk atmospheric composition, provided the atmosphere is 
relatively clear of hazes or clouds. In the presence of extended vertical hazes, the transmission spectrum will instead 
bear the signature of the hazes' optical properties (\citealt{H_B_2012}; \citealt{Robinson_2014}; \citealt{W_S_2015}).
Either way, this slope provides an important constraint on the properties of exoplanetary atmospheres, and has been 
detected in a few exoplanets, e.g. HD~209458b (\citealt{Knutson_2007}; \citealt{Sing_2008}), HD~189733b \citep{Pont_2008}, 
HAT-P-5b \citep{Southworth_2012}, WASP-12b \citep{Sing_2013}, GJ~3470b \citep{Nascimbeni_2013},
WASP-6b (\citealt{Jordan_2013}; \citealt{Nikolov_2015}),
Qatar-2b \citep{Mancini_2014}, 
WASP-31b \citep{Sing_2015}, and WASP-103b \citep{Southworth_2015}.

In spectral region where molecular absorption dominates, the atmospheric scaleheight can still be estimated from 
the size of absorption features if one assumes that molecules are well-mixed in the atmosphere. However, it is 
very difficult to determine the absolute abundances of species and the pressure regions that are probed, as degenerate 
solutions exist which can match observations. Indeed, changes of several orders of magnitude in molecular abundances 
can be compensated with a small change in the planetary radius at the reference pressure \citep{Griffith_2014}. Although 
the absolute mole fraction of species is ambiguous, their relative abundances are well constrained through
the relative strength of absorption features \citep{B_S_2012} in a haze-free atmosphere. 

However, none of these previous investigations of the retrieval of atmospheric parameters considers the refractive 
effects of the exoplanetary atmosphere on its transmission spectrum, in stark contrast with stellar occultations. 
The use of refractive effects to probe planetary atmospheres has a long standing history in Solar system exploration
which started with the occultation of $\sigma$~Arietis by the atmosphere of Jupiter \citep{B_C_1953}. This pioneering 
work paved the way for stellar occultation by planetary atmospheres to become a powerful method by which one can derive 
not only the density vertical profile of planetary atmospheres, but also deduce the vertical profile of minor chemical 
species which absorb stellar radiation (see the review by \citealt{S_H_1990}). However, the observational geometry of a 
stellar occultation is different from that of a transiting exoplanet. Indeed, in a stellar 
occultation, the radiation source is infinitely far away compared to the planet and the observer, 
whereas in an exoplanet transit geometry the observer is infinitely far away from the 
exoplanet and its star. 

\citet{Hubbard2001} first investigated the 
impact of refraction on exoplanet transmission spectra and deduced that refractive effects were negligible 
for the hot Jupiter HD~209458b. However, \citet{H_S_2002} adapted gravitational microlensing theory to atmospheric 
refraction and showed that refraction could be important for some transiting exoplanets. \citet{S_S_2010} used a 
more conventional ray-tracing approach and came to the same conclusion. They both found that even in the absence 
of absorption, an exoplanet's effective radius is still quite large because the higher density regions of 
planetary atmospheres deflect stellar radiation away from the observer. Absorption by chemical species were 
later combined with ray-tracing (\citealt{GarciaMunoz_2012}; \citealt{YB_LK_2013}; Misra, Meadows \& Crisp 2014) 
to compute transmission spectra of an Earth-Sun analog, and showed that refraction decreases 
significantly the contrast of H$_2$O features because transmission spectroscopy cannot probe most of the 
troposphere, which holds the bulk of Earth's water inventory.

\citet{YB_LK_2014} explored the parameter space of an Earth-like planet receiving the 
same stellar flux as Venus, Earth, and Mars, for different spectral type host stars along the Main 
Sequence. They showed that one can only probe an Earth-like planet with the same insolation as Earth 
to a pressure of 1~bar if the planet orbits a M5 or cooler star, assuming it maintains a similar 
temperature-pressure profile. As the temperature of the host star increases, the contrast of spectral 
features decreases, potentially resulting in flat spectral regions (see their Figure~11).
Refraction also produces wings brighter than the continuum just outside 
of transit (\citealt{H_S_2002}; \citealt{S_S_2010}; \citealt{GarciaMunoz_2012}) because some of the 
stellar radiation is deflected toward the observer while the star is unocculted. These brighter wings 
are sensitive to the aerosol content of an atmosphere \citep{M_M_2014}.

More recently, \citet{YB_LK_2015} developed new analytical expressions to describe the integrated column 
abundance along the curved path of a refracted ray, as well as the total deflection of the ray. They
discovered that thick atmospheres form a thin refractive boundary layer where the 
column abundance and its deflection, go from finite to infinite 
values as the grazing altitude of a ray decreases and approaches a lower refractive boundary. Indeed, at this lower 
boundary, the radius of curvature of the ray induced by refraction matches the radial position at which the ray 
grazes the atmosphere, and the ray follows a closed circular path. These results diverge substantially from expressions 
developed for stellar occultations (Baum \& Code 1953) and still in use today. Rays that penetrate below this lower 
boundary will spiral into the atmosphere and be absorbed. Hence, atmospheric regions located below this lower boundary 
cannot be probed by techniques based on the transmission of radiation through an atmosphere, irrespective of the 
planet-star geometry. Furthermore, the effective scaleheight of the atmosphere, i.e. as perceived by an observer, 
decreases dramatically as this lower boundary is approached.

What impact does refraction, and in particular this new theory, has on transmission spectroscopy of giant exoplanets? 
Does it affect the Rayleigh scattering slope, or the retrieval of atmospheric parameters? This paper explores these
issues for Jupiter-sized planets with a Jovian composition across various planetary temperatures and spectral type of 
the host star.

\section{Details of simulations}\label{simul}

To illustrate the impact of refraction on transmission spectroscopy of giant planets and on the retrieval of their atmospheric 
properties, I compute a suite of transmission spectra from 0.4 to 5.0~$\micron$ for a Jovian-like planet with a H$_2$-He atmosphere 
orbiting stars of different spectral type along the Main Sequence (M2, K5, G2, F0). I consider isothermal spherically-symmetric 
well-mixed atmospheres for three different temperatures (300, 600, and 1200~K), and various constant mole fractions of water vapour
($0$, $10^{-12}$, $10^{-10}$, $10^{-8}$, $10^{-6}$, and $10^{-4}$). These simulations are done both with and without 
refraction. 
Stellar limb-darkening is ignored in these simulations, just as it was ignored in all 
derivations of the Rayleigh slope (\citealt{Lecavelier_2008}; 
\citealt{B_S_2012}; \citealt{H_B_2012};
\citealt{dW_S_2013}), to have a common reference frame
and enable a direct comparison with previous simulations of 
transit spectra.
Much of the theory and methods at the heart of these simulations have already been described in detail 
(\citealt{YB_LK_2013}, 2014, 2015), but are summarised in this section for the convenience of the readers.

\subsection{Planetary atmospheres}\label{atmos}

I build-up the density 
vertical profile of an isothermal homegeneous atmospheres in hydrostatic equilibrium with a gravitational acceleration 
which varies with altitude, for three different temperatures (300, 600, and 1200~K), in the manner described 
in Section~3.3 in \citet{YB_LK_2015}. The radius ($R_P=69911$~km) and mass ($M_P=1.8986~\times~10^{27}$~kg) of the 
exoplanet are the same as Jupiter's. The atmosphere has a He mole fraction of 0.1357, as measured on Jupiter by the 
Galileo probe (listed in \citealt{L_F_1998}), while the remainder of the atmosphere is predominantly composed 
of H$_2$ and trace amounts of water. The planetary radius ($R_p$) is defined at a reference pressure of 1~atm (atmosphere), 
and the highest pressure considered in the simulations, or ``surface'' pressure, is 500~atm. 
The densities are computed on a fine computational grid with thickness $\Delta z_{comp}$, and then sampled on a 
coarser grid with altitude intervals $\Delta z_{samp}$. Table~\ref{table1} lists the thickness of the layers making the
computational and sampling grids, along with the chosen atmospheric thickness ($\Delta Z_{atm}$), and the resulting 
``surface'' altitude ($z_s$) for each temperature. The atmospheric thickness defines the radius of the top atmospheric 
boundary ($R_{top}$), given by
\begin{equation}
 R_{top} = R_p + z_s + \Delta Z_{atm} .
\end{equation}

\subsection{Planet-star distance}\label{dist}

To determine the distance between the planet and its stellar host, I assume that the isothermal temperature of 
the planet ($T_p$) matches the mean planetary emission temperature and that the planet has a circular orbit. 
Under these assumptions, the semi-major axis ($a$) of the planet is given by,
\begin{equation}\label{equte}
 a = \frac{\sqrt{1 - \Lambda_B}}{2} \left(\frac{T_*}{T_p} \right)^{2} R_*
\end{equation}
where $\Lambda_B$ is the Bond albedo of the planet, while $T_*$ and $R_*$ are the stellar effective temperature 
and radius, respectively. Table~\ref{table2} lists these stellar parameters for the four stellar spectral type 
(M2, K5, G2, F0) along the Main Sequence considered in this paper. A Bond albedo of 0.30 is chosen, similar to
those of the giant planets in our Solar system.

\subsection{Column abundances and ray deflection}\label{abundef}

To compute the transmission spectrum of an exoplanet, one must determine the altitude-dependent incremental column abundance 
encountered by a ray along its curved refracted path, as well as the total deflection of the ray, as a function of a 
ray's grazing height ($r_{0}$), i.e. planetary radius~+~grazing altitude of the ray. 
I compute these quantities with MAKEXOSHELL (see Section~3 in \citealt{YB_LK_2015}), including the contribution from atmospheric 
regions above the top atmospheric boundary, with 80 rays spread uniformly in grazing altitude between the top and bottom
boundaries. These grazing altitudes also define the boundaries of the atmospheric layers used in the computation. 
For the refractive case, I use a refractivity at standard temperature and pressure ($\nu_{STP}$) of 1.23$\times$10$^{-4}$, that of 
the Jovian atmosphere at $1~\micron$, and a refractivity of 0 otherwise. The location of the bottom atmospheric boundary 
depends on the condition of the simulations. In the refractive case, the bottom boundary can be either the critical or the 
lower refractive boundary (see Section~\ref{boundaries}), otherwise the bottom boundary is the ``surface''. 

\begin{table}
 \centering
 \begin{minipage}{60mm}
  \caption{Input and derived model atmosphere parameters for different atmospheric temperatures}
  \begin{tabular}{@{}lccc@{}}
  \hline
 \multicolumn{4}{c}{Input}  \\
 \hline
 $T_p$ (K) & 300 & 600 & 1200 \\
 $\Delta$Z$_{atm}$ (km) & 1370 & 2760 & 5640 \\
 $\Delta$z$_{samp}$ (km) & 5 & 10 & 20 \\
 $\Delta$z$_{comp}$ (km) & 0.01 & 0.02 & 0.04 \\ 
\hline
 \multicolumn{4}{c}{Derived}      \\
\hline
 $z_{s}$ (km) & -260.70 & -519.48 & -1031.28 \\
\hline
\end{tabular}\label{table1}
\end{minipage}
\end{table}

\begin{table}
 \centering
 \begin{minipage}{52mm}
  \caption{Stellar parameters \citep{cox2000}} 
  \begin{tabular}{ccc}
 \hline
 Spectral type & $T_*$ (K) & $R_*$ ($R_\odot$) \\
\hline
 F0 & 7300 & 1.50  \\
 G2 & 5778\footnote{\citet{L_F_1998}} & 1.00 \\ 
 K5 & 4410 & 0.72  \\
 M2\footnote{values from \citet{R_H_2005}} & 3400 & 0.44  \\ 
\hline
\end{tabular}\label{table2}
\end{minipage}
\end{table}

\begin{table*}
 \centering
 \begin{minipage}{120mm}
  \caption{Refractive boundaries of Jupiter-sized planets with different isothermal temperatures orbiting stars of 
various spectral type (see Section~\ref{boundaries})}
  \begin{tabular}{@{}ccccccc@{}}
  \hline
 T (K)  & \multicolumn{2}{c}{300} & \multicolumn{2}{c}{600} & \multicolumn{2}{c}{1200} \\
 Critical Boundary & n (amagat\footnote{1 amagat = density at standard temperature and pressure (STP)}) & P (bar) & n (amagat) & P (bar) & n (amagat) & P (bar) \\
 \hline
 F0 star & 0.33 & 0.37 & 1.78 & 3.97 & 8.64 & 38.45  \\
 G2 star & 0.54 & 0.60 & 2.78 & 6.20 & 12.21 & 54.34  \\ 
 K5 star & 0.92 & 1.02 & 4.45 & 9.90 & 16.25 & 72.35  \\
 M2 star & 1.56 & 1.73 & 6.62 & 14.74 & 18.75 & 83.45 \\ 
\hline
 Lower Boundary \footnote{independent of stellar spectral type} &  4.88  &  5.43  &  9.74  &  21.69  &  19.38 &  86.27 \\
\hline
\end{tabular}\label{table3}
\end{minipage}
\end{table*}

\subsection{Critical and lower refractive boundaries}\label{boundaries}

The lower refractive boundary is located at a height ($r$) where, in a spherically symmetric atmosphere, 
a grazing ray follows a circular path (see Section~2.2 in \citealt{YB_LK_2015}). Atmospheric regions 
below this boundary cannot be probed by exoplanet transmission spectroscopy or by stellar occultations because rays 
that reach these regions spiral deeper into the atmosphere until they are absorbed. This boundary is located 
at a height where
\begin{equation}\label{condition}
\left( \frac{\nu(r)}{1+\nu(r)} \right) \frac{r}{H} = 1  ,
\end{equation}
and this depends only on the properties of the atmosphere.
Here, $\nu(r)$ is the height-dependent refractivity of the atmosphere, and $H$ is the density scale height. 

In reality, one can only probe to the critical refractive boundary (see Section~2.2 in \citealt{YB_LK_2014}). 
This boundary occurs at a density where the atmosphere deflects a ray coming from the opposite stellar limb 
toward the observer. Hence, this boundary depends both on the properties of the atmosphere and on 
the geometry of the planet-star system, i.e. on the angular size of the star as seen from the planet.
When the planet occults the central region of its host star, this boundary is located at the same density
all around the planetary limb, where rays are deflected by a critical deflection ($\omega_c$). During the course 
of a transit, this symmetry is broken and one can probe higher 
densities on one side of the planet, and lower densities on the opposite side. However, since the ray deflection 
tends to infinity as the grazing height of the ray approaches the lower boundary, the critical boundary can only 
asymptotically approach but never reach the lower boundary.

MAKEXOSHELL computes the quantity on the left-hand side of equation~\ref{condition} on the fine altitude 
computational grid, and determines the lower boundary by choosing the altitude for which this value 
is closest to but less than one. I also use MAKEXOSHELL to find the location of the critical boundary 
by computing the ray deflection for a grid of grazing heights, and then determine which one has the smallest
difference with the critical deflection of the planet-star system, which is given by
\begin{equation} \label{critdef}
\sin\omega_{c} = \frac{r_0 + R_{\ast}}{a} .
\end{equation}
Table~\ref{table3} shows the critical boundaries for the four spectral type stars from Table~\ref{table2}, 
both in terms of the density and pressure of the atmosphere, for the three isothermal temperatures 
considered. It also lists the lower boundaries which are independent of the spectral type of the host star.

\subsection{Radiative transfer}

The radiative transfer code \citep{YB_LK_2013} combines the MAKEXOSHELL-computed column abundances ($N_{ik}$) 
of the $k^{th}$ atmospheric layer encountered by the $i^{th}$ ray, with the average mole fraction ($f_{ijk}$) and 
extinction cross section ($\sigma_j$) of the $j^{th}$ chemical species, to compute the integrated 
optical depth ($\tau_i$) along the $i^{th}$ ray, using
\begin{equation}
\tau_i = \sum_{k} \overline{\sigma}_{ik} N_{ik} ,
\end{equation}
where 
\begin{equation}
 \overline{\sigma}_{ik} = \sum_{j} \sigma_j f_{ijk}
\end{equation}
is the average extinction cross section of the $k^{th}$ layer along the $i^{th}$ ray. Since all the 
molecular species are well-mixed in these simulations, $f_{ijk}$ is independent of the atmospheric layer 
and of the ray, and is simply the average mole fraction of the $j^{th}$ species. The transmission ($T_i$)
of the  $i^{th}$ ray is simply given by 
\begin{equation}
 T_i = e^{- \tau_i}  .
\end{equation}
Earlier versions of the code (see e.g. \citealt{Johnson_1995}; \citealt{K_T_2009}) 
added thermal emission to the transmitted stellar radiation. However, for a one-dimensional spherically
symmetric atmosphere, the thermal emission contribution to an exoplanet light curve is constant throughout
the transit, and the calculated transit depth is independent of thermal emission. To save 
on computation time, \citet{YB_LK_2013} modified the code so that one can choose not to include thermal 
emission.

Although the line opacity database is identical to that in \citet{K_T_2009}, \citet{YB_LK_2013} replaced the original 
Rayleigh scattering parametrisation, appropriate only for Earth, with a database of Rayleigh scattering cross 
sections for N$_2$, O$_2$, CO$_2$, and Ar (see their Table~2). They also included new routines to compute 
optical depths from continuous absorption cross sections, and a new database of ultraviolet and visible 
absorbers (see their Table~1). 

For this paper, I have added the Rayleigh scattering cross section 
of molecular hydrogen and helium to the database. I use the formulation by \citet{F_B_1973} for the 
H$_2$ cross section as a function of its rotational quantum number J, from which I compute the cross section 
for an equilibrium H$_2$ population at 300~K. Since the temperature dependence of the cross section is very 
small (e.g. about 0.6 per cent change between 273.15 and 2000~K at 1~\micron), I use the 300~K cross section for 
all temperatures. For helium, I compute its STP refractivity with the expression from \citet{M_P_1969}, also
described in \citet{Weber_2003}, and obtain its Rayleigh cross section ($\sigma_R$) with Rayleigh's formula:
\begin{equation}
 \sigma_R = \frac{32\pi^{3}}{3} \left( \frac{{\nu_{STP}}}{n_{STP}} \right)^{2} w^{4} F_K . 
\end{equation}
Here, $n_{STP}$ is the number density at standard temperature and pressure, also known as 
Loschmidt's number, $w$ is the wavenumber of the radiation, and $F_K$ is the King correction 
factor. For helium, $F_K$ is equal to one at all wavenumbers.

The only other source of opacity considered in these simulations is that of water vapour. 
The algorithm for the line-by-line computation of optical depths is described 
in \citet{T_S_1976}. To reduce the computation time, only H$_2$O lines which have a peak optical 
depth greater than $10^{-7}$ within an atmospheric layer are included, and the line profile is 
computed to an optical depth of $10^{-8}$ or to the edge of the spectral region, whichever is 
closer to the centre of the line.
The atmospheric transmission is computed from 1800 to 26000~cm$^{-1}$ every 0.05~cm$^{-1}$, and 
then averaged in 4~cm$^{-1}$-wide bins. I choose a spectral region which is larger than that of 
interest to mitigate edge effects. The transmission spectrum is eventually displayed
with respect to wavelength from 0.4 to 5.0~\micron.

To compute the effective radius ($R_{eff}$) of the transiting exoplanet, I use the 80 rays
from MAKEXOSHELL (see Section~\ref{abundef}) to partition the atmosphere into 80 annuli 
evenly spread in altitude. I then map these altitude boundaries to their corresponding impact 
parameter ($b_i$) across the planetary limb, which is what an observer perceives.
The impact parameter is given by
\begin{equation}
 b_i = ( 1 + \nu(r_{0i}) ) r_{0i} .
\end{equation} 
Taking the transmission of each annulus to be the average of the transmission at the boundaries, and ignoring stellar limb-darkening, the 
effective radius is then given by
\begin{equation}
 R_{eff}^{2} = R_{top}^{2} - \sum_{i = 1}^{N} \left( \frac{T_{i+1} + T_{i}}{2} \right) (b_{i+1}^{2} - b_{i}^{2}) \label{reff}
\end{equation}
\citep{YB_LK_2013} where $N$ is the number of annuli. The $(N+1)^{th}$ ray is that which grazes the top atmospheric boundary, 
chosen so that both atmospheric absorption and deflection are negligible. Thus $T_{N+1} = 1$, and $b_{N+1} = R_{top}$. 
Atmospheric regions with densities higher than the density at the lower atmospheric boundary (i.e. the ``surface'', lower refractive 
boundary, or critical refractive boundary depending on the simulation) are implicitly assumed to be 
opaque at all wavelengths. One advantage to including refraction is that one does not need to compute the 
transmission of atmospheric regions located at higher densities and pressures, or lower altitude, than the lower 
refractive boundary. I display the transmission spectrum in terms of the effective atmospheric thickness ($\Delta z_{eff}$) of the 
transiting exoplanet with respect to the 500~atm ``surface'', which I compute with
\begin{equation}
 \Delta z_{eff} = R_{eff} - (R_p + z_s) .
\end{equation}

\section{Results and discussion}\label{results}

\subsection{Rayleigh scattering slope}\label{raylslope}

\begin{figure}
\includegraphics[scale=0.55]{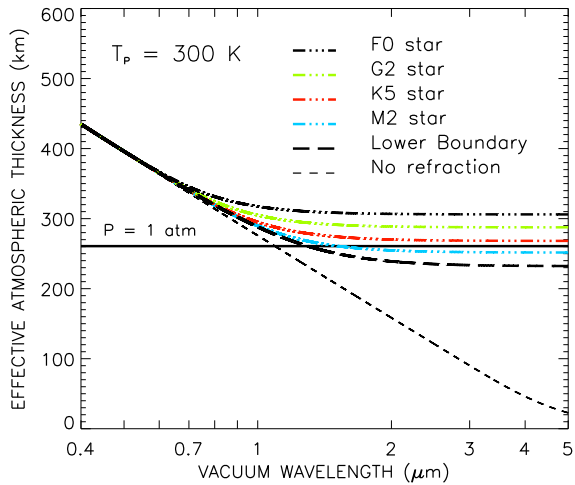}
\caption{H$_2$-He Rayleigh scattering transmission spectra of a Jupiter-sized exoplanet with 
a 300~K isothermal Jovian atmosphere, expressed in term of the atmosphere's effective thickness 
above the 500~atm pressure level (as for all figures). The short-dashed line shows the spectrum without 
refraction, while the coloured triple-dot-dashed lines show spectra with refraction
for different Main Sequence spectral type of the host star. The long-dashed line
shows the spectrum with refraction if one could somehow probe the atmosphere to the lower refractive
boundary, irrespective of the planet-star distance. The thick horizontal solid line shows the altitude 
of the 1~atm pressure level, for reference (shown in all figures). See Table~\ref{table3} for the density 
and pressure levels of the critical and lower refractive boundaries. 
\label{fig1}}
\end{figure}

\begin{figure}
\includegraphics[scale=0.55]{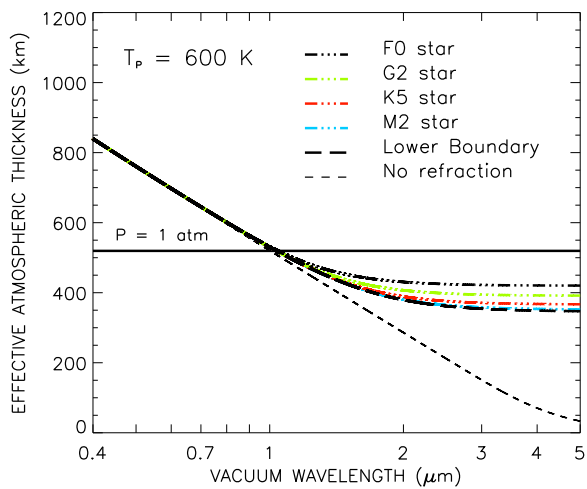}
\caption{H$_2$-He Rayleigh scattering transmission spectra of a Jupiter-sized exoplanet with
a 600~K isothermal jovian atmosphere. See caption in Fig.~\ref{fig1} for more details.
\label{fig2}}
\end{figure}

\begin{figure}
\includegraphics[scale=0.55]{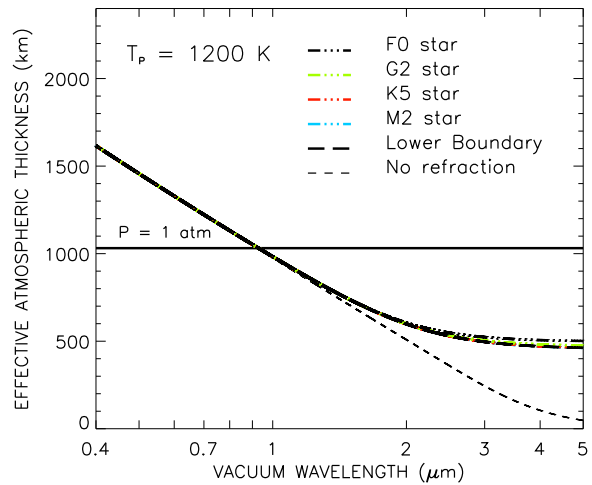}
\caption{H$_2$-He Rayleigh scattering transmission spectra of a Jupiter-sized exoplanet with
a 1200~K isothermal jovian atmosphere. See caption in Fig.~\ref{fig1} for more details.
\label{fig3}}
\end{figure}

To explore how refraction affects the Rayleigh scattering slope, I first compute transmission spectra
where the only source of opacity is Rayleigh scattering from H$_2$ and He. 
Figures~\ref{fig1} through~\ref{fig3} show the 0.4-5.0~$\micron$ H$_2$-He Rayleigh continuum
under various conditions for the 300, 600, and 1200~K isothermal atmospheres, respectively.
The legends in all three figures are identical. The 500~atm ``surface'' is located at 0~km altitude, 
and the 1~atm pressure level is indicated for reference. Since the main effect of temperature is 
to change the scale height of the atmosphere, I scale the range of values on the ordinate with the 
planetary temperature so that approximately the same number of scaleheight are displayed in all figures. 
It is not exactly the same because the gravitational acceleration varies with altitude in the simulations, 
and so does the scaleheight. It is useful to plot things this way to notice secondary effects subtler than the primary 
effect of the scaling of spectral features with scaleheight. Indeed, even if transmission spectra have the same 
aspect from one figure to the next, the actual size of spectral features still differ by roughly the ratio of 
the planetary temperatures. I have done this in all figures throughout the paper.

Without refraction, the transmission spectrum of a Rayleigh scattering isothermal atmosphere should 
follow a straight line when plotted in terms of the effective atmospheric thickness versus the natural
logarithm of the wavelength (\citealt{Lecavelier_2008}; \citealt{B_S_2012}; \citealt{H_B_2012}), as long as
the Rayleigh scattering cross section follows a constant power law with wavelength. The Rayleigh continuum
for the non-refractive case (shown by the short-dashed line in all three figures) has roughly a 
constant slope except longward of 3~$\micron$ where the Rayleigh cross section is sufficiently weak that the 
opaque regions below the ``surface'' contribute significantly to the transmission spectrum. In this case, 
these regions introduce a significant step function in the altitude-dependent atmospheric transmission profile, thus 
violating the conditions under which the result from \citet{Lecavelier_2008} holds, and changing the Rayleigh 
slope.

A noticeable secondary effect is that the spectrum also shifts to lower altitudes with increasing temperatures. 
This occurs because I reference the planetary radius to a pressure, as is often done (e.g. \citealt{B_S_2012}; 
Swain, Line, \& Deroo 2014), rather than to a density. As temperature increases, the same pressure is 
achieved with lower densities. Thus the densities required to achieve a slant optical thickness 
of 0.56 are located at lower altitudes and the transmission spectrum has an extra downward shift, 
beyond the primary scaleheight-induced factor, as the planetary temperature increases. It would seem that 
referencing the planetary radius to a density would be a better choice as the transmission spectrum 
would not shift up or down with temperature, but I nevertheless adopt this common convention in this paper.

In the refractive case, this simple picture changes dramatically. Indeed, the black long-dashed
line shows the transmission spectrum if one can somehow probe to the lower refractive boundary, irrespective 
of the planet-star distance, which produces the smallest differences between the refractive and non-refractive
cases. The infrared transmission spectrum is flat, 
similar to that of an optically thick 
cloud or a surface, and gradually merges with the non-refractive case at lower wavelengths. The wavelength
above which the Rayleigh slope deviates significantly from the non-refractive case increases with temperature, 
and changes from about 0.6~$\micron$ at 300~K to about 1~$\micron$ at 1200~K. The fractional change 
of the spectrum with and without refraction increases with wavelength and decreases with temperature,
and it is still quite important even at 1200~K above 3~$\micron$.

Accounting for the location of the critical boundary associated with the spectral type of the host star 
increases the effective thickness of the atmosphere and changes the Rayleigh slope further (see triple-dotted 
dashed coloured lines in Figures~\ref{fig1} through \ref{fig3}). This correction can be quite important 
at the cooler planetary temperatures (see Figure~\ref{fig1}). As the temperature of the planet increases, 
the differences between the spectra associated with a F0 and a M2 star decrease. At 1200~K (see Figure~\ref{fig3}), 
there is hardly any differences between the transmission spectrum around a F0 and a M2 star, and both are 
very similar to the spectrum where one can probe to the lower refractive boundary. Indeed, as the planetary 
temperature increases, the critical boundary approaches the lower refractive boundary. Since the effective 
atmospheric scaleheight decreases dramatically at the approach of the lower refractive boundary (see 
Section~4.2.3 in \citealt{YB_LK_2015}), the altitude separation between the different critical 
boundaries also decreases.

Hence, the discovery of the refractive boundary layer by \citet{YB_LK_2015} turns out to be 
extremely important for the transmission spectra of giant exoplanets, because even in the absence of clouds or hazes,
one cannot probe the atmosphere to arbitrarily large pressures in spectral regions between molecular bands.
Indeed, this refractive boundary layer not only mimics a surface but also decreases substantially the 
effective scaleheight of the atmosphere near that boundary. This, in turns, modifies not only the effective 
radius of the Rayleigh continuum of the hotter exoplanets, but also decreases the Rayleigh scattering slope of the 
transmission spectrum. If one wants to infer the atmospheric scaleheight \citep{Lecavelier_2008} or the mass 
of the exoplanet \citep{dW_S_2013} from transmission spectra, one must be careful to evaluate the Rayleigh 
scattering slope in spectral regions that are largely unaffected by refraction, given the planetary temperatures. 
However, even then, the Rayleigh slope might still be affected by clouds or hazes, depending on their 
vertical distribution. 

\begin{figure}
\includegraphics[scale=0.55]{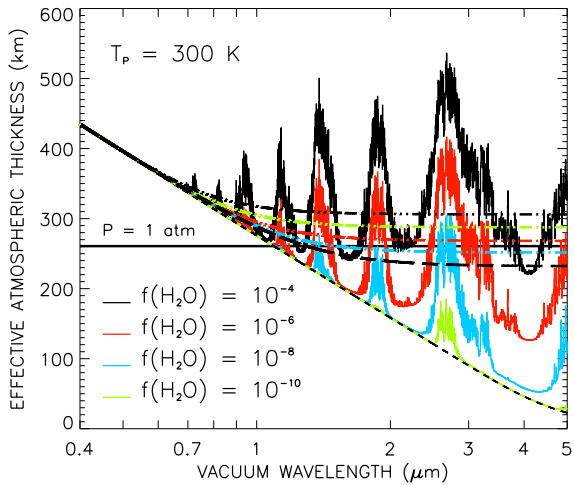}
\caption{Transmission spectra without refraction of a Jupiter-sized exoplanet with a 300~K isothermal Jovian atmosphere
for various vertical homogeneous mole fractions of water (see inset legend). 
The water-free Rayleigh scattering transmission spectra from Fig.~\ref{fig1} are overlaid for comparison (see 
caption and inset legend in Fig.~\ref{fig1} for more details). Refraction will suppress water features which lie below 
the refractive Rayleigh curve associated with the host star's spectral type (coloured triple-dotted dashed lines). 
\label{fig4}}
\end{figure}

\begin{figure}
\includegraphics[scale=0.55]{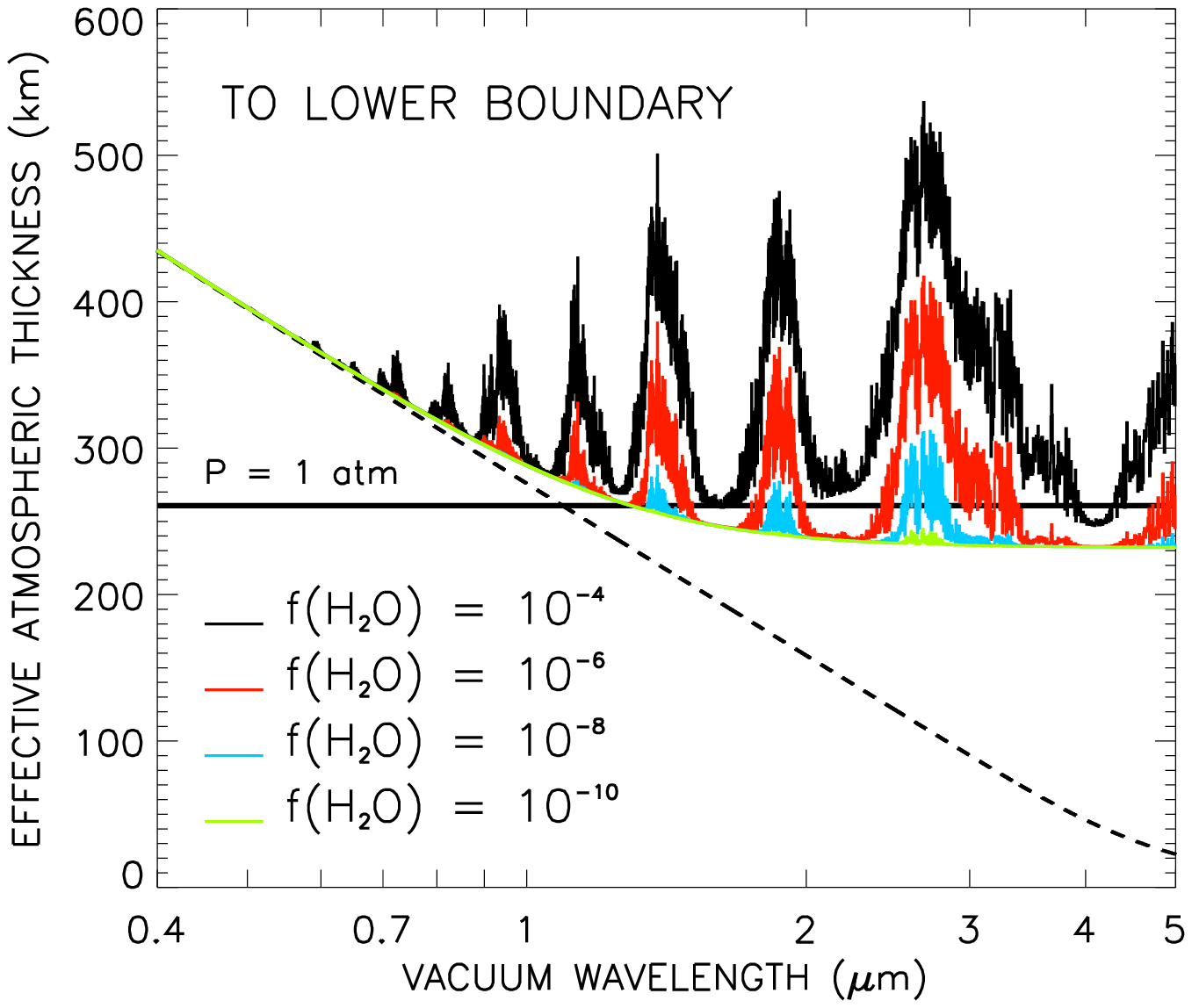}
\caption{Transmission spectra with refraction of a Jupiter-sized exoplanet, with a 300~K isothermal Jovian atmosphere
for various vertical homogeneous mole fractions of water (see inset legend), if one could 
somehow probe the atmosphere to the lower refractive boundary, irrespective of the planet-star distance.
The non-refractive Rayleigh continuum (short-dashed curve) is shown for comparison.
\label{fig5}}
\end{figure}

\begin{figure}
\includegraphics[scale=0.55]{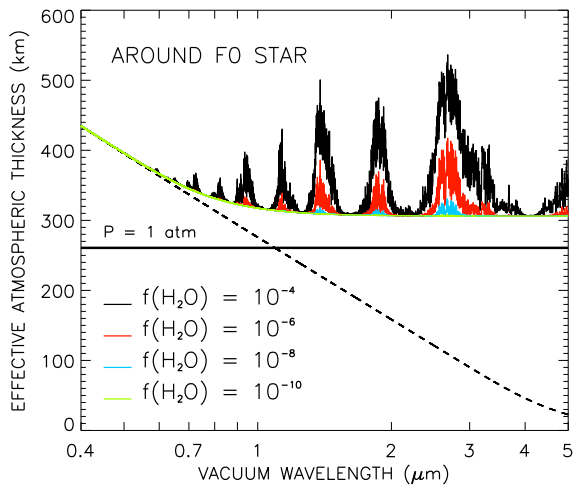}
\caption{Transmission spectra with refraction of a Jupiter-sized exoplanet, with a 300~K isothermal Jovian atmosphere
for various vertical homogeneous mole fractions of water (see inset legend), orbiting a F0~star.
The non-refractive Rayleigh continuum (short-dashed curve) is shown for comparison.
\label{fig6}}
\end{figure}

\subsection{Abundance retrieval}

To explore how refraction affects the retrieval of atmospheric abundances of chemical species, I now add 
various mole fractions of vertically-mixed water vapour ($10^{-12}$, $10^{-10}$, $10^{-8}$, $10^{-6}$, and $10^{-4}$)
to the simulations. I use water vapour as an example because its spectral signature is often sought after in 
exoplanetary transmission spectra, but the discussion in this section is generally applicable to other 
molecules (e.g. CH$_4$, NH$_3$, CO, CO$_2$) as well.

Figure~\ref{fig4} shows the resulting transmission spectra (coloured solid lines) in the non-refractive case 
for the 300~K isothermal atmosphere compared with the various Rayleigh continua from Figure~\ref{fig1}. 
Note that a H$_2$O mole fraction of $10^{-12}$ does not produce any noticeable absorption features above the Rayleigh continuum 
and is not shown. Changing the water abundance mostly shifts features up or down with respect to the non-refractive Rayleigh 
continuum (short-dashed line).
In spectral region where molecular absorption is not much stronger than Rayleigh scattering, i.e. from 0.7 to 2~$\micron$ for 
H$_2$O, the size of absorption features are pretty sensitive to the abundance of the responsible species. However, with the 
rapid decrease with wavelength of the Rayleigh scattering cross section, water abundances can change by 
several orders of magnitude above a certain threshold with hardly a change in the size of the spectral features
in the infrared. In this regime, small changes in the retrieved planetary radius can compensate for several orders of 
magnitude of change in the retrieved abundance \citep{Griffith_2014}, thus resulting in certain retrieval degeneracies 
when faced with noisy data or data obtained only in the infrared.

\begin{figure}
\includegraphics[scale=0.55]{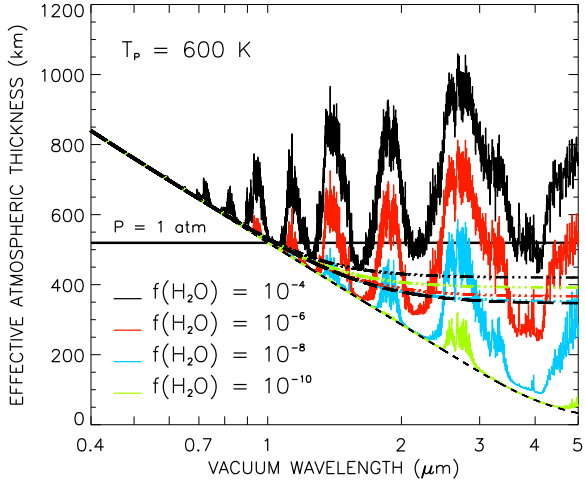}
\caption{Transmission spectra without refraction of a Jupiter-sized exoplanet with a 600~K isothermal Jovian atmosphere
for various vertical homogeneous mole fractions of water (see inset legend). 
The water-free Rayleigh scattering transmission spectra from Fig.~\ref{fig2} are overlaid for comparison (see 
caption and inset legend in Fig.~\ref{fig1} for more details). Refraction will suppress water features which lie below 
the refractive Rayleigh curve associated with the host star's spectral type (coloured triple-dotted dashed lines).
\label{fig7}}
\end{figure}

\begin{figure}
\includegraphics[scale=0.55]{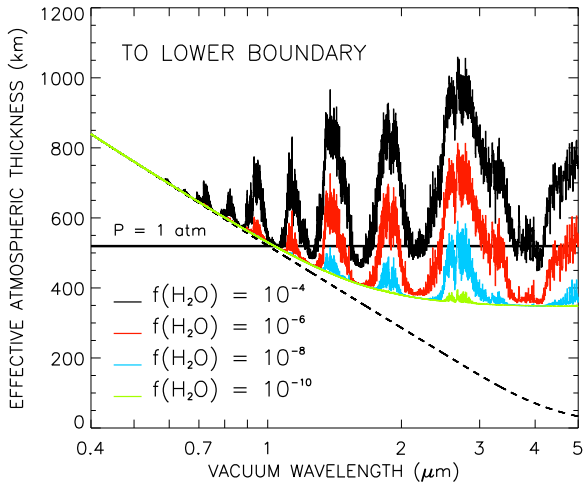}
\caption{Transmission spectra with refraction of a Jupiter-sized exoplanet, with a 600~K isothermal Jovian atmosphere
for various vertical homogeneous mole fractions of water (see inset legend), if one could 
somehow probe the atmosphere to the lower refractive boundary, irrespective of the planet-star distance.
The non-refractive Rayleigh continuum (short-dashed curve) is shown for comparison.
\label{fig8}}
\end{figure}

Figure~\ref{fig4} also shows that much of the water features for the non-refractive case can lie below the 
refractive Rayleigh continuum, depending on the water abundance and the spectral type of the host star. As these features will be 
severely decreased in the refractive case, Figure~\ref{fig4} illustrates the fraction of water features that can be hidden by 
refraction for various scenarios. In the refractive case, the size of absorption features increases as the critical boundary 
decreases in altitude. The maximum size occurs if one can somehow probe down to the lower refractive boundary, 
which is shown in Figure~\ref{fig5}. Since the effective scaleheight of the atmosphere tends to zero toward the lower refractive 
boundary, refraction reduces the size of spectral features the closer they are to this boundary. This is particularly obvious with 
the peak of the water feature at 2.7~$\micron$ for various abundances. Without refraction (Figure~\ref{fig4}), 
the size of this feature for a mole fraction of $10^{-10}$ (green solid line), is similar to the part of the feature above 
the long-dashed solid line for a mole fraction of $10^{-8}$ (blue solid line).
However, their sizes in the refractive case (Figure~\ref{fig5}) are completely different. 

\begin{figure}
\includegraphics[scale=0.55]{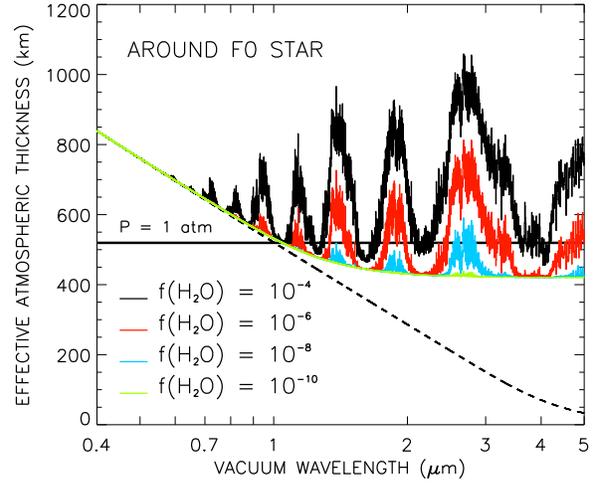}
\caption{Transmission spectra with refraction of a Jupiter-sized exoplanet, with a 600~K isothermal Jovian atmosphere
for various vertical homogeneous mole fractions of water (see inset legend), orbiting a F0~star.
The non-refractive Rayleigh continuum (short-dashed curve) is shown for comparison.
\label{fig9}}
\end{figure}

What is particularly interesting about Figure~\ref{fig5} is that, unlike in the non-refractive case, the size of spectral features 
in the infrared varies dramatically with the abundance of the absorbing species. Hence, the retrieval degeneracy described by 
\citet{Griffith_2014} does not exist for a cloud-free 300~K cool atmosphere because the refractive lower boundary provides
a reference altitude akin to a surface below which spectral features cannot exist. Furthermore, the boundary which is relevant for 
the retrieval of atmospheric properties is the critical boundary, which raises the refractive
continuum and decreases the size of 
spectral features further, and more so for hotter stars than for cooler stars. Amongst the spectral type that I have considered
(see Table~\ref{table2}), a planet orbiting an F0 star has the smallest spectral features. Figure~\ref{fig6} shows that around a F0 star
a 300~K atmosphere with a water mole fraction of $10^{-10}$ does not have any recognisable features, and an abundance of 
$10^{-8}$ produces only weak features, contrary to what is observed for the non-refractive case (Figure~\ref{fig4}).

\begin{figure}
\includegraphics[scale=0.55]{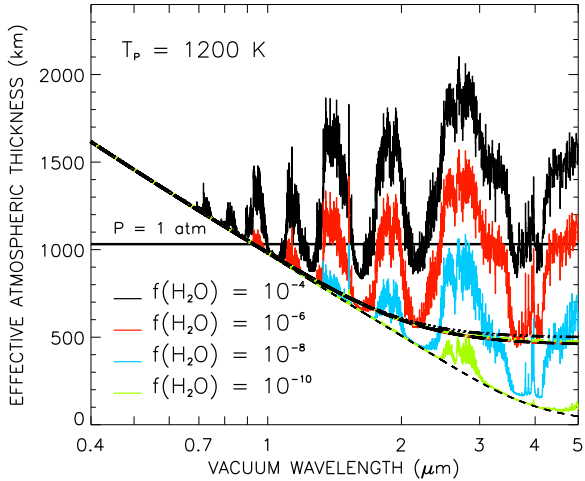}
\caption{Transmission spectra without refraction of a Jupiter-sized exoplanet with a 1200~K isothermal 
jovian atmosphere for various vertical homogeneous mole fractions of water (see inset legend). 
The water-free Rayleigh scattering transmission spectra from Fig.~\ref{fig3} are overlaid for comparison (see 
caption and inset legend in Fig.~\ref{fig1} for more details). Refraction will suppress water features which lie below 
the refractive Rayleigh curve associated with the host star's spectral type (coloured triple-dotted dashed lines).
\label{fig10}}
\end{figure}

\begin{figure}
\includegraphics[scale=0.55]{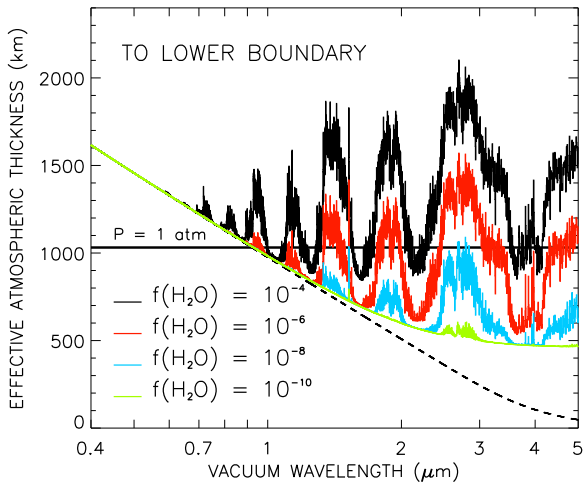}
\caption{Transmission spectra with refraction of a Jupiter-sized exoplanet, with a 1200~K isothermal jovian atmosphere
for various vertical homogeneous mole fractions of water (see inset legend), if one could 
somehow probe the atmosphere to the lower refractive boundary, irrespective of the planet-star distance.
The non-refractive Rayleigh continuum (short-dashed curve) is shown for comparison.
\label{fig11}}
\end{figure}

The breaking of this degeneracy is not 
due to opacity from Rayleigh scattering but rather by two 
refractive effects also responsible for the decrease
of the Rayleigh slope:
\begin{itemize}
\item{The decrease of the effective scaleheight of the atmosphere toward a value of zero as rays approach the 
lower boundary.}
\item{The deflection, away from the oberver, of rays grazing atmospheric regions below the critical altitude. 
This creates a step function with altitude 
in the perceived illumination of the atmosphere, 
just as a surface or an optically thick cloud does.}
\end{itemize}
These two effects create a refractive continuum which manifests itself in the infrared, where Rayleigh 
scatering is weak, in spectral regions of low molecular opacity.

\begin{figure}
\includegraphics[scale=0.55]{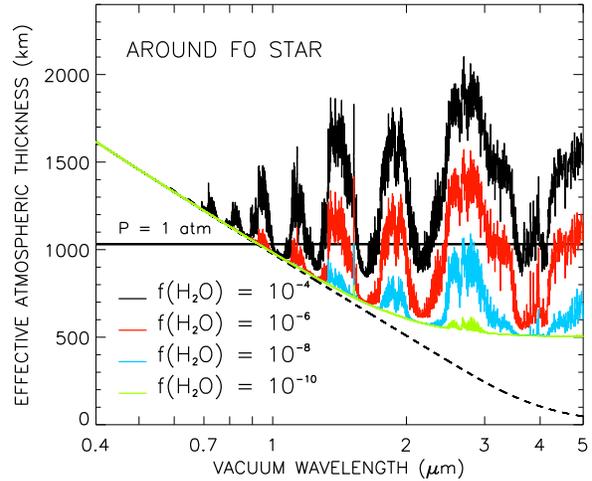}
\caption{Transmission spectra with refraction of a Jupiter-sized exoplanet, with a 1200~K isothermal jovian atmosphere
for various vertical homogeneous mole fractions of water (see inset legend), orbiting a F0~star.
The non-refractive Rayleigh continuum (short-dashed curve) is shown for comparison.
\label{fig12}}
\end{figure}

Figures~\ref{fig7}--\ref{fig9}, and Figures~\ref{fig10}--\ref{fig12} display the same information as Figures~\ref{fig4}--\ref{fig6}, 
but for planetary temperatures of 600 and 1200~K, respectively. Aside from the scaling of the size of the spectral features with 
temperature, one can observe only subtle changes with temperature in the overall shape of the spectral features in the non-refractive case
(see Figures~\ref{fig4}, \ref{fig7}, and \ref{fig10}). In the refractive case, as the planetary temperature increases, differences
in the spectral type of the host star becomes less important and are almost negligible at 1200~K (compare Figures~\ref{fig11} 
and \ref{fig12}). Also, the size of spectral features
varies with abundances but these variations are not as pronounced for the hotter exoplanets as the cooler exoplanets (e.g. compare 
Figures~\ref{fig5}, \ref{fig8}, and \ref{fig11} when one can probe to the lower boundary). This occurs because, as the planetary 
temperature increases, the refractive continuum 
is lower relative to the non-refractive H$_2$O spectral 
features, and thus has a correspondingly lower impact on the shape of these features. 

At temperatures of 1200~K, a water mole fraction of $10^{-6}$ or higher produces spectral features
which are largely unaffected by refraction, a result that seems consistent with \citet{Hubbard2001}. However, including 
refraction allows one to constrain the possible range of abundances better. If one observes strong features 
in the data then small abundances are excluded. Conversely, the non-detection of spectral features need no longer be 
explained only with clouds, but can also with low abundances.

What happens to inferred abundances if an optically thick cloud deck is located at a lower density or pressure, or higher altitude,
than the critical boundary? In this case, the cloud deck will raise the continuum and cut down the size of spectral features even further
than refraction does, so that the spectral features might have the same size as if produced by a lower abundance in a clear atmosphere. Hence, if one assumes that the atmosphere
is cloud-free, abundances inferred from transmission 
spectra will be lower limits.

\section{Conclusions}

I investigated the effects of refraction on transmission spectra of giant exoplanets with a suite of 
simple isothermal atmospheric models for a Jupiter-like exoplanet with a Jovian H$_2$-He composition,
ignoring stellar limb-darkening. 
The effects are particularly important for giant exoplanets, more so than for terrestrial exoplanets, 
because they do not possess a surface which could otherwise limit the density, or pressure, to which one 
can probe their atmosphere. 

Indeed, without refraction, one can probe a pure Jovian H$_2$-He Rayleigh scattering 
atmosphere to more than 500~bar of pressure at 5~$\micron$, barring other sources of extinction. 
However, refraction creates a thin refractive boundary layer at high densities where the perceived 
density scaleheight of the atmosphere is dramatically smaller than the actual scaleheight \citep{YB_LK_2015}. 
This apparent decrease in the scaleheight reduces the size of spectral features and decreases the Rayleigh 
scattering slope within that layer. 

The lower boundary of this refractive boundary layer is defined where the refraction-induced curvature of 
a ray allows it to follow a circular trajectory around the planet. Atmospheric regions below this boundary 
can never be probed during an exoplanetary transit. This flattens out the Rayleigh continuum in the infrared 
and effectively mimics a surface. The pressure level of this lower refractive boundary increases with temperature
and is much lower than in the non-refractive case. It varies from a pressure of 5.4 to 86.3~bar as the atmospheric 
temperature changes from 300 to 1200~K in an isothermal Jovian atmosphere.

The flat infrared Rayleigh continuum that one obtains with refraction is significantly different from the non-refractive case, 
where the Rayleigh continuum decreases with wavelength with a slope characteristic of the scaleheight of the atmosphere. 
At shorter wavelengths, the refractive Rayleigh continuum slowly merges with the non-refractive one. The wavelength above 
which refraction changes significantly the Rayleigh slope increases with temperature, from about 0.6 to 1~$\micron$ 
between atmospheric temperatures of 300 and 1200~K. Accounting for the angular size of the host star seen from the planet, 
raises the Rayleigh continuum and changes the Rayleigh slope further, which shifts this spectral boundary to lower wavelengths. 

The magnitude of this correction can be quite substantial for an atmospheric temperature of 300~K and for a F0 spectral
type host star, but decreases with decreasing stellar temperature and with increasing atmospheric temperature.
At atmospheric temperatures of 1200~K, the correction for a planet around an F0 star is very small. At this temperature, 
the transmission spectrum is almost independent of the spectral type of the host star and is mostly determined 
by the location of the lower refractive boundary.

As I illustrated with water absorption features, 
the reduced effective scaleheight near the lower 
boundary and the step function in the apparent 
illumination of the atmosphere at the critical boundary, 
also responsible for the reduction of the Rayleigh slope, 
break the retrieval degeneracy between abundances and 
planetary radius. In the non-refractive case, when 
absorption is much 
stronger than Rayleigh scattering, changes of several orders of magnitude in the mole fraction of well-mixed species 
only shifts the spectrum up or down in effective radius without changing the size of spectral features significantly. 
However, in the refractive case, the size of H$_2$O features changes dramatically with abundances for 
a 300~K atmosphere. As the atmospheric temperature increases, the refractive continuum shifts downward relative 
to spectral features, and the variation with abundance of the size of spectral features decreases. At atmospheric 
temperatures of 1200~K, only water mole fraction lower than $10^{-6}$ show some significant size variation with abundance.

The possible presence of an optically thick cloud deck complicates retrieval of atmospheric properties further. 
The abundances obtained from assuming a clear atmosphere are lower limits. However, the brighter refractive wings 
just outside of transit can be used to distinguish between clear and cloudy atmospheres \citep{M_M_2014}. Hence, it is 
clear that including all effects of refraction in exoplanet atmosphere retrieval algorithms can only help constrain 
atmospheric properties.

Although quite important even for hot exoplanets, refractive effects are more pronounced in cooler atmospheres. 
Even though the bulk of exoplanet observations has been primarily focused on hot exoplanets because they are easier to detect 
and characterise, long-time monitoring of various areas of the sky is just starting to yield longer-period cooler exoplanets 
(\citealt{Crossfield_2015}, \citealt{Jenkins_2015}) with promises of more to come. Furthermore, with the upcoming planned 
launches of the Transiting Exoplanet Survey Satellite (TESS) and the James Webb Space Telescope (JWST), it is only a matter of 
time before we start obtaining exoplanet transmission spectra of sufficient quality that signatures of refraction can be 
exploited to constrain the atmospheric properties of planets of any temperature.

\section*{Acknowledgements}

I wish to thank the directors of the Max-Planck-Institut f\"ur Astronomie, Thomas Henning and 
Hans-Walter Rix, who have graciously provided a fantastic research environment and facilities for 
one full year above and beyond the end of my contract. This research, as well as that in
\citet{YB_LK_2015}, would not have been possible otherwise.

\label{lastpage}

\end{document}